# HTTPI BASED WEB SERVICE SECURITY OVER SOAP

Pankaj Choudhary   Rajendra Aaseri   Nirmal Roberts

ABV - Indian Institute of Information Technology & Management
Gwalior, India

`pschoudhary007@gmail.com`, `jmsrdjbr@gmail.com`, `nirmal@iiitm.ac.in`

## ABSTRACT

*Now a days, a new family of web applications 'open applications', are emerging (e.g., Social Networking, News and Blogging). Generally, these open applications are non-confidential. The security needs of these applications are only client/server authentication and data integrity. For securing these open applications, effectively and efficiently, HTTPI, a new transport protocol is proposed, which ensures the entire security requirements of open applications. Benefit of using the HTTPI is that it is economical in use, well-suited for cache proxies, like HTTP is, and provides security against many Internet attacks (Server Impersonation and Message Modification) like HTTPS does. In terms of performance HTTPI is very close to the HTTP, but much better than HTTPS. A Web service is a method of communication between two ends over the Internet. These web services are developed over XML and HTTP. Today, most of the open applications use web services for most of their operations. For securing these web services, security design based on HTTPI is proposed. Our work involves securing the web services over SOAP, based on the HTTPI. This secure web service might be applicable for open applications, where authentication and integrity is needed, but no confidentiality required.*

*In our paper, we introduce a web service security model based on HTTPI protocol over SOAP and develop a preliminary implementation of this model. We also analyze the performance of our approach through an experiment and show that our proposed approach provides higher throughput, lower average response time and lower response size than HTTPS based web service security approach.*

## 1. INTRODUCTION

HTTP [4] and HTTPS [5] are two most accepted transport protocols used on web. HTTP does not provide any security assurance, but it is flexible, lightweight and supported by cache proxies in Internet. On the other hand, HTTPS (HTTP over TLS/SSL) provides all three security assurances: server authentication, integrity and confidentiality; however HTTPS is less flexible, heavyweight, has no support for cache proxies and also additional latency in network. Hence, HTTPS is avoided by many websites and web applications. Several web applications are rising in recent times, which are non-confidential and require only authentication and message integrity as their security needs. These applications are called open applications [12]. Examples of such open applications are Social Networking, News and Blogging applications. HTTPI [1], [2], a new protocol was proposed for these open applications, with two security guarantees, data integrity and client/server authentication, but no guarantee of any data confidentiality. Benefit of using HTTPI in place of HTTPS is support by cache proxies and security against many cyber attacks like Server Impersonation and Message Modification. It is also mentioned in [1], [2] and [3] that the HTTPI has nearly same performance as the HTTP and much better performance than HTTPS. Thus the HTTPI is lightweight, effective and inexpensive protocol for securing open applications. Web services are platform-independent and autonomous services, which can be easily programmed, described, discovered and published over the network. These applications can be web-based, local, or distributed [9]. Web services are developed on the top of open standards such as HTTP, Java, HTML, and XML [15]. To secure these web services, WS-Security [6], [7], [8], defines the use of SOAP extensions [16] to apply client/server authentication, integrity and confidentiality, for securing the Web Service on the message level. WS-Security makes use of the XML security, like Binary Security Tokens and XML-Digital Signatures, as the building block. The majority of non-confidential open applications, like Social Networking, Blogging applications and News sites, make use of web services for most of their operations. It is found that using HTTPS for providing security to open applications adds significant overhead and also lack in performance [1], [2]. For secure and efficient deployment of these open web services, a web service security model which is based on the design concept of the HTTPI protocol can be devised.

In this paper we present the idea of securing web service over SOAP, based on the HTTPI protocol design. Our security goal is to provide the client/server authentication and message integrity for our Web Service. We do not require the message confidentiality. In this context for authentication we are using Username/Password Tokens and Binary Authentication Tokens (X.509 certificates) [18]. For Message Integrity XML Digital Signature [17] is used with RSA-SHA1 signature algorithm. Also for secure communication between two web services a Non-Encrypted session is set-up. This HTTPI based secured web service can be used in open applications in future to secure them effectively and efficiently in terms of authentication and integrity.

In particular, the following tasks will be performed:

- An examination of web services and various web services security scenarios.
- Design model of a web service security over SOAP, depending on the design of HTTPI protocol.
- Build up a prototypical implementation of the design.
- Performance evaluation of our implementation by means of a web application with other scenarios.

## 2. RECENT WORK

T. Choi et al. proposed an idea and a design concept of HTTPI [1]. Again, they proposed the protocol HTTPI [2]. Its security is comparable to HTTPS, while its performance and scalability is comparable to HTTP. The protocol ensures authentication and integrity, but does not provide message confidentiality.

Singh K. et al. proposed a concept of HTTPi for web content integrity [3] that uses the concept of SHTTP, which offers end-to-end message integrity and no confidentiality. It is easy to set up for today's web architecture and has efficient design.

Lindstrom, P. et al. and C. Geuer et al. presented a report on Attacking and Defending Web Services and web service security standards [10], [11]. In these reports they provided various types of threats, attacks and proposed defending techniques against these attacks on Web Services.

Nadalin, A. et al. proposed the Web Services Security: SOAP Message Security 1.0, [7] and Hirsch, F. et al. proposed the Web Services Security: SOAP Messages with Attachments 1.1, OASIS Standard [6]. These specifications explain improvement to SOAP messaging to offer authentication, message integrity and a broad range of security models.

## 3. PROPOSED SCHEME: WEB SERVICE SECURITY BASED ON HTTPI

For designing the web service security based on the HTTPI protocol we need to use WS-Security for applying the authentication, integrity at the message level and also for managing the non-encrypted session for secure communication between two services, as there is no need of the confidentiality in HTTPI design. WS-Security describes a specific set of SOAP extensions for development of secure web services. In Figure we show the use of WS-Security elements for providing authentication by means of security tokens and integrity by means of signature element and timestamp element. Here, the data in SOAP body is non-confidential.

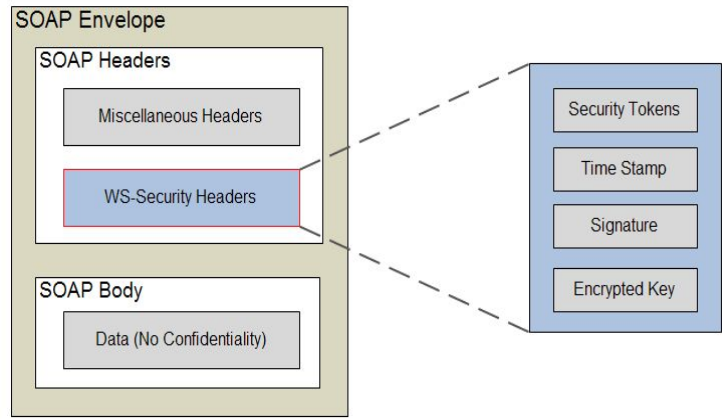

Figure 1. WS-Security for SOAP Message (Authentication and Integrity)

### 3.1. Authentication

In our approach for providing authentication to our web service, we used Username/Password Tokens and X.509 certificates as Binary Authentication Tokens.

i.      *Username/Password Authentication:*

In WS-Security <UsernameToken> element is used, for providing custom authentication using username and password validation.

```
<UsernameToken>
  <Username>User_1</Username>
  <Password
Type="PasswordDigest">fmdsf7sM0RpIhBQFgfsdfSjdim/asdj=</Password>
  <Nonce>dfEdf2ywdsfkMDx5sfEsfzWw==</Nonce>
  <Created>2012-12-12T12:35:45Z</Created>
</UsernameToken>
```

In above example, the password is the digest of the real text password, a Nonce and the security token creation time, so,

*PasswordDigest = Base64(SHA-1(Password + Nonce + Created))*

Here, a nonce random string (uniquely generated), that is used to identify the password and to avoid the replay attack.

**ii.**     *Binary authentication tokens in the form X.509 certificates:*

For adding and transmitting an X.509 certificate [18] with a SOAP message, *<BinarySecurityToken>* element is used, which contains the certificate's public version. The certificate itself is sent encoded as base64.

```
<wsse:BinarySecurityToken
  ValueType="wsse:X509v3"
  EncodingType="wsse:Base64Binary"
  Id="SecurityToken-c4sf8j-236e-9fd3-3688-3847df8d30s8">
  sfhutzCCFDKfKAwIkriyQ83hyHlslmd/opBsfdapmcbkjK...
</wsse:BinarySecurityToken>
```

In above example a X.509 certificate is passed in *<BinarySecurityToken>* element.

## 3.2. Integrity

For providing Message Integrity to our web service, we use XML Digital Signature [17] and RSA-SHA1 as signature algorithm.

XML signatures are digital signatures intended to apply in XML communication to add integrity, authentication, and maintain the non-repudiation of signed data [21]. In WS-Security for using digital signatures, <Signature> element is used as part of a security header with some sub-elements.

The fundamental composition of XML Digital Signature is as follows:

```
<Signature>
  <SignedInfo>
    <CanonicalizationMethod />
    <SignatureMethod />
    <Reference>
       <Transforms>
       <DigestMethod>
       <DigestValue>
    </Reference>
    <Reference />
  </SignedInfo>
  <SignatureValue />
  <KeyInfo />
</Signature>
```

- **SignedInfo:** having the signed data or reference of it and also specifies algorithms used.
- **SignatureMethod:** specifies algorithms used for signature.
- **Reference:** define the resources which are to be signed included in URI reference.

- **Transforms:** specifies, if any transforms to apply to the resource before signing.
- **DigestMethod:** defines the hash/digest algorithm for hashing.
- **DigestValue:** contains the value after the hash algorithm is applied to the resources.
- **SignatureValue:** holds the outcome of the Base64 encoded signature, which is the actually generated signature.
- **KeyInfo:** provides the key (generally as X.509 digital certificates) to recipients for validation of the signature.

### 3.3. Session Management

By setting up a session context between two web services, they can communicate securely. This is done by using security extensions to SOAP for session management, e.g., the <Continue> element [13], [14]. The <Continue> element is just like 'handshake' and it is to initiate the session and to agree on the way they will communicate. The <Continue> holds a <Nonce> element, an blank <Session/> element and a <Nr/> element set to 1. The <Nonce> element by the Client web service (cws) is a randomly generated number. This Nonce is added to identify the serving web service (sws). The <Nr> element indicates how many SOAP message the each web service has sent within this session.

In most cases for open applications, the web services only need to communicate in such a way that, messages are authenticated and integrity has been preserved. In such cases the message encryption is not needed. For such cases Non-Encrypted session setup is done.

**Non-Encrypted Session setup:**

This non-encrypted session has the series of three SOAP messages sent forward and backward. The reason to set up this session is to make sure that the two web services are securely communicating to each other.

The steps to setup the non-encrypted session are following:

i. The client web service (cws) A sends a SOAP envelope with a digital signature using its own private key and a <Continue> element with a sub element as the last element in the SOAP message body.

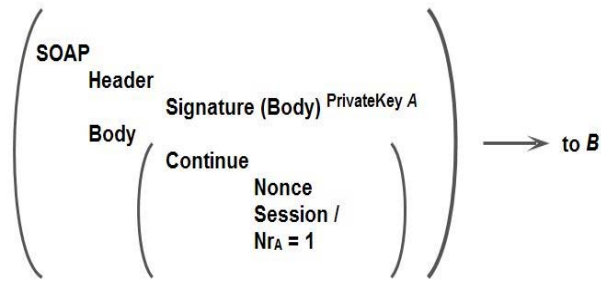

Figure 2. SOAP message from cws A to sws B

ii. When message is received, the serving web service (sws) B, decides if it is in agreement with the method of communication. In return, the sws send a SOAP envelope back with the digital signature of the body. The <Continue> holds three sub elements, the <Nonce> element it received, the session element packed with session id and the <Nr> element. The session id is concatenation of a timestamp and a randomly generated number to ensure uniqueness.

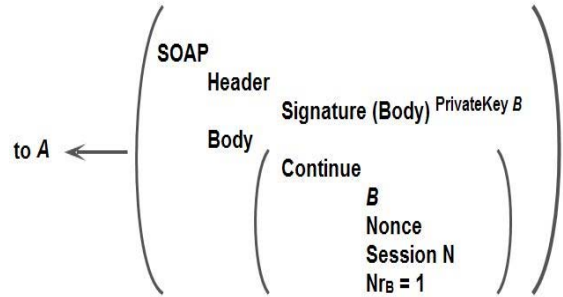

Figure 3. SOAP message from sws B to cws A

iii.       Again, the client web service (cws) sends a SOAP envelope with a digital signature using its own private key and the request message for operation to be performed with a <Continue> element with two sub elements, a session id and <Nr> element in the SOAP message body. After this step the Non-Encrypted session is established between client and serving web services. For ending the session any web service can send a <SessionEnd/> element with a SOAP envelope.

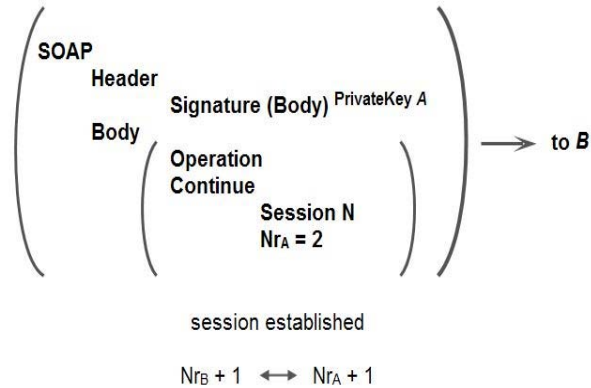

Figure 4. Non-encrypted session established between cws A and sws B

The above figure represents the order in which SOAP messages are interchanged between client web service A and serving web service B. With above session setup, both web services can describe the operations of each other and also exchange information in secure manner and thus no opponent can obstruct them.

## 4. EXPERIMENT & RESULTS

In this section we explain the experiment, in which we invoke a sample web service, and secure it based on the HTTPI design idea, by using two open source web service tools, soapUI (as a web service client) and WSO2 Application Server (as a web service server). SoapUI [19] is an openly available web service application for testing service-oriented architectures (SOA), generally used for web service development, invoking, functional and load testing and simulation. WSO2 Application Server [20] is a Web services engine by Apache Axis2. It is a trivial, high performing tool for Web Services, and offers a safe, operational and trustworthy runtime for installing and supervising Web services.

### 4.1. Experimental Methodology and Setup

For applying security on Web Services, we write the WS-Policy document as per our particular security requirement and then engage that policy into our Web Service [22]. For HTTPI based security scenario, we employ a policy which implements the following constraints:

- Applying the Asymmetric Binding.
- Containing the Time Stamp in request.
- Use of RSA-SHA1 signature algorithm for signing SOAP body.
- Use of Basic-256 algorithmic suit.

In our experiment, we configure our web service over four different security scenarios (including our HTTPI based security scenario) at WSO2 Application Server and soapUI.

**i.**     *No Security Scenario:*

In this scenario, we deploy a SOAP based Web Service in WSO2 Application Server and invoke it with soapUI. There is no security policy applied yet to this Web Service.

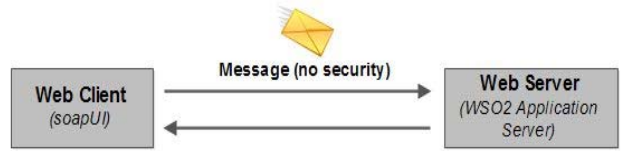

Figure 5. No-Security Scenario

**ii.**     *Username/Password Scenario:*

In this scenario, Username token is used in our deployed Web Service for authentication. At transport level, web client sends the Username Token inside message, and web server validates it by finding it in user store of service's end. This scenario provides authentication to client.

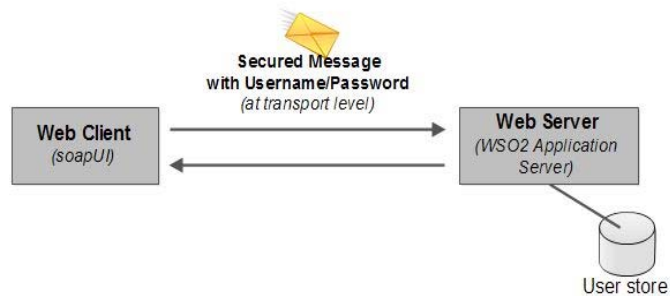

Figure 6. Username/Password Scenario

**iii.**     *HTTPI Based Scenario:*

In this scenario, both web client and web server have X.509 certificates and Keystores. The client's Keystore have private key of client and public certificate of server. The server's Keystore have private key of server and public certificate of client. The request and response messages of Web Service are signed by using the private key of the sender and verified by using public key of signing party. The scenario is HTTPI based and it provides the Authentication and Integrity to our deployed Web Service.

61

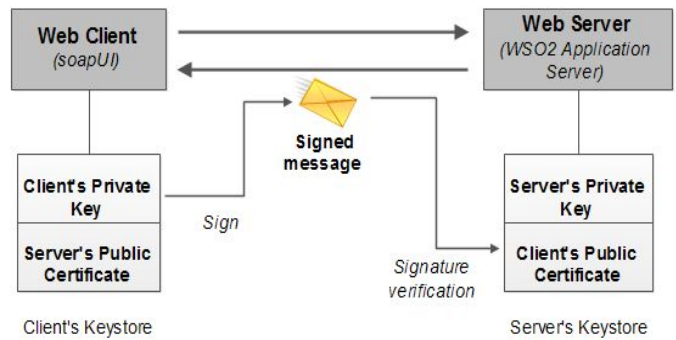

Figure 7. HTTPI Based Scenario

iv.    *Signature & Encryption Scenario (HTTPS Based Scenario):*

In this scenario, both web client and server have X.509 certificate and their own Keystores. As in previous scenario the client's Keystore and server's Keystore contains the same keys and certificates. The request and response messages of Web Service are encrypted by using public key of receiver and signed using private key of sender. At other end, these messages are decrypted using private key of receiver and message signatures are validated using public key of sender. This scenario HTTPS based and provides all three security guarantees: Authentication, Integrity and Confidentiality.

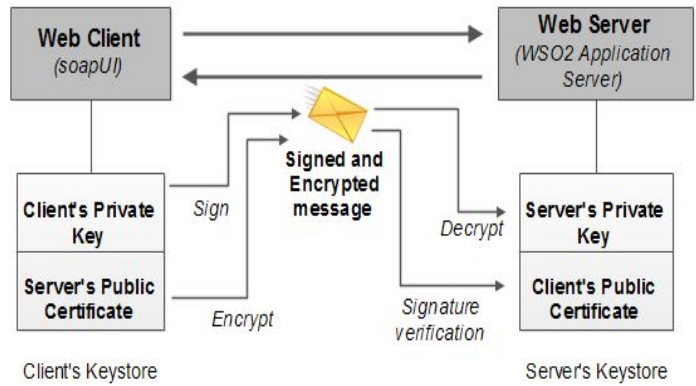

Figure 8. HTTPS Based Scenario

## 4.2. Experimental Results

After configuring all these four security scenarios on the soapUI and WSO2 Application Server, now we conducted a performance test to analyze the performance of our HTTPI based web service security scenario with HTTPS based and other security scenarios, in terms of throughput (transaction/seconds), average response time (milliseconds), and response size (KB). For performance testing, we used soapUI load testing feature [19], with various number of SOAP requests for deployed Web Service, configured for all four scenarios.

Table 1. Performance Metrics obtained for Web Service using soapUI

| Scenarios | Average Response Time (milliseconds) | Average Throughput (transaction/second) | Reply Size per Request (Bytes) |
|---|---|---|---|
| No Security Scenario | 19.56 | 6.816 | 276 |
| Username/Password Scenario | 26.20 | 5.790 | 713 |
| HTTPI Based Scenario | 30.45 | 5.148 | 864 |
| Signature and Encryption Scenario | 43.168 | 4.016 | 1964 |

In this experiment, we took number of virtual users = 5, and number of requests = 10, 20, 40, 60 up to 500. We ran each stage for each set of number of requests, for all four scenarios, and measured three parameters: throughput (transaction/seconds), average response time (milliseconds), and response size (KB). The results obtained from our experiment are shown in figure 9(a) to figure 9(c).

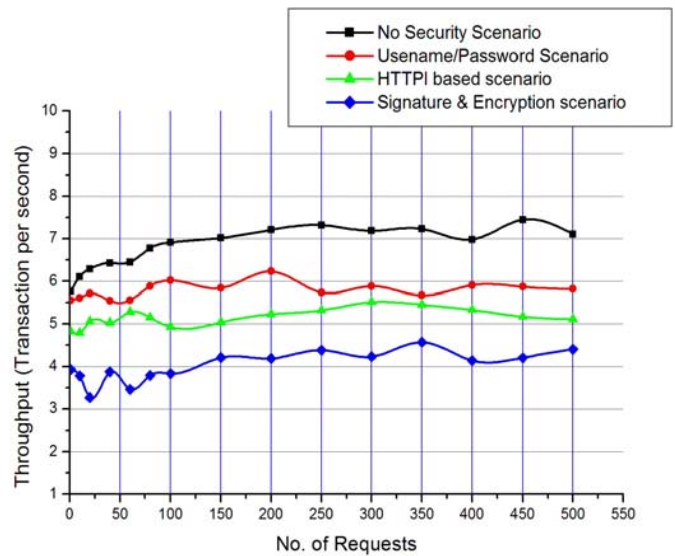

(a) Throughput vs. No. of Requests

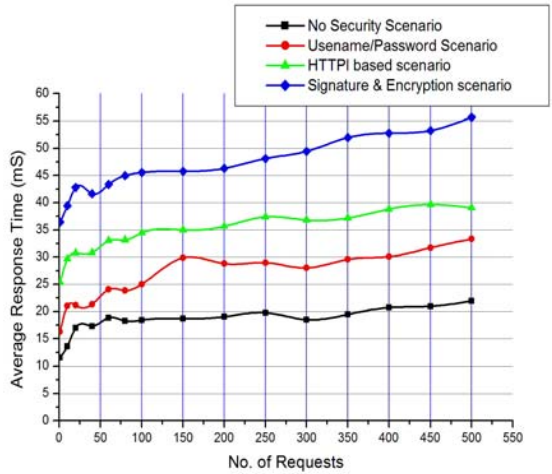

(b) Average Response Time vs. No. of Requests

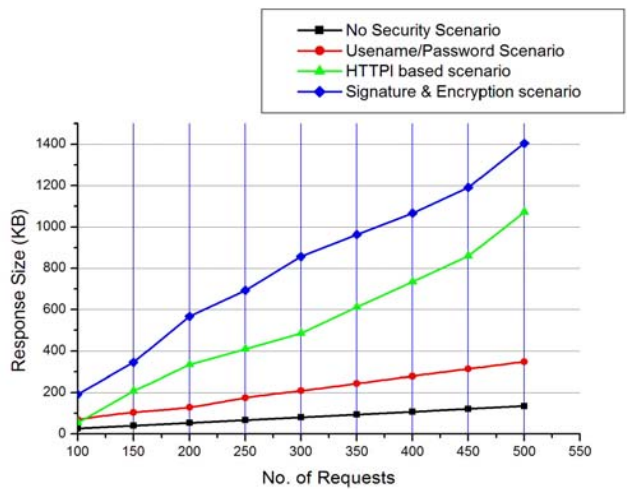

(c) Response Size vs. No. of Requests

Figure 9. Performance Comparison for all four scenarios

For entirety, we repeated our experiment with some other Web Services for all four scenarios, configured them in same way, with some more number of virtual users and requests. We compared the performance again on same parameters. The results show the same tendency as previously shown in above figures.

## 5. CONCLUSION AND FUTURE SCOPE

In this paper, we propose a web service security model based on HTTPI protocol over SOAP, with the security goal: client/server authentication and integrity on message, without confidentiality. As per our proposed scheme, we used Username/Password Tokens and Binary Authentication Tokens (X.509 certificates) for Authentication and XML Digital Signature (with RSA-SHA1 as a signature algorithm) for Message Integrity. We set up a Non-Encrypted session to secure the communication between two web services.

We examined the performance of our scheme through an experiment. From the results of our experiment, we conclude that our HTTPI based web service security scenario provides higher throughput (in transaction/seconds), lower average response time (in milliseconds), and lower response size (in KB) than HTTPS based web service security scenario, when there is no need of message confidentiality, and having little overhead over the Non-secured and Username/Password scenario.

Thus, the secured web services based on HTTPI can be used in non-confidential open applications (like: Social Networking, Blogging and News sites) in future to secure them effectively and efficiently in terms of authentication and integrity.